\documentclass[conference]{IEEEtran}
\IEEEoverridecommandlockouts
\usepackage{cite}
\usepackage{algorithm}
\usepackage{algpseudocode}
\usepackage{amsmath}
\usepackage{amsfonts}
\usepackage{amsmath,amssymb,amsfonts}
\usepackage{bm}
\usepackage{enumerate}
\usepackage{subfigure}
\usepackage{graphicx}
\usepackage{textcomp}
\newtheorem{theorem} {Theorem}
\newtheorem{remark}[theorem]{Remark}
\def\BibTeX{{\rm B\kern-.05em{\sc i\kern-.025em b}\kern-.08em
    T\kern-.1667em\lower.7ex\hbox{E}\kern-.125emX}}
\begin{document}

\title{A Novel Random Access Scheme for
M2M Communication in Crowded
Asynchronous Massive MIMO Systems}

\author{Huimei Han$^{1}$, Wenchao Zhai$^2$, Zhefu Wu$^1$, Ying Li$^3$,  Jun Zhao$^4$, Mingda Chen$^1$, 
\vspace{1.5mm}
\\
\fontsize{10}{10}\selectfont\itshape
$^1$College of Information Engineering, Zhejiang University of Technology,
Hangzhou, Zhejiang Province, China\\
$^2$College of Information Engineering, China Jiliang University,
Hangzhou, Zhejiang Province, China\\
$^3$State Key Lab of Integrated Services Networks, Xidian University, Xi¡¯an,
710071, P.R. China\\
$^4$School of Computer Science and Engineering, Nanyang Technological University, Singapore\\
\vspace{1.5mm}
\\
\fontsize{9}{9}
$^1$\{hmhan1215,\,wzf,\,2111803126\}@zjut.edu.cn, $^2$ zhaiwenchao@cjlu.edu.cn, $^3$yli@mail.xidian.edu.cn, $^4${junzhao}@ntu.edu.sg \vspace{-2mm}
}

\maketitle

\begin{abstract}
A new random access scheme is proposed to solve the intra-cell pilot collision for M2M communication in crowded asynchronous massive multiple-input multiple-output (MIMO) systems. The proposed scheme utilizes the proposed estimation of signal parameters via rotational invariance technique enhanced (ESPRIT-E) method to estimate the  effective timing offsets, and then active UEs obtain their timing errors from the effective timing offsets for uplink message transmission.  We analyze the  mean squared error   of the  estimated effective timing offsets of UEs, and the  uplink throughput. Simulation results show that, compared to the exiting random access scheme for the crowded asynchronous massive MIMO systems, the proposed scheme can improve the uplink throughput and estimate the effective timing offsets  accurately at the same time.
\end{abstract}


\section{Introduction}\label{sec:introduction}
 The machine-to-machine (M2M) communication is centered on the intelligent interaction of  user equipment (UEs) without human intervention,  which is the enabler for the Internet of Things (IoT) to achieve the envision of the ``Internet of Everything"~\cite{iot1,iot2}. In recent years, the M2M communication developments rapidly and  has been applied to many scenarios, such as smart medical, smart vehicle, smart logistics, etc. Cisco visual networking index and forecast  predicts that there will be around 28.5 billion connected UEs by 2022~\cite{crowdue}. The massive multiple-input multiple-output (MIMO) technology, which achieves significant improvements in  energy  and spectral efficiency and serve massive UEs in the same time-frequency resource, is well suited for the  M2M communication~\cite{effi1,nora}.

For the M2M communication in massive MIMO systems, the number of UEs in the cell is envisioned in the order of hundreds or thousands, and the payload
data generated by the M2M traffic is usually in small size~\cite{mMTC1}. Random access  procedure  is the first step to initiate a data transmission, which is an important  step  in the M2M communication systems~\cite{randomaccess}. The connection-oriented random access procedure utilized in the long term evolution (LTE) network may induce excessive signaling overhead and cannot support massive access~\cite{randomaccess}.

Researchers are exploring new random access schemes for M2M communication in massive MIMO systems, and the grant-based random access schemes  have been proposed in recent years. E.~{Bj\"ornson}~\textit{et~al.}~proposed a strongest-user collision resolution (SUCRe) scheme, which allocates the pilot to the  UE with largest channel gain among the contenders~\cite{SUCR}. However, the number of successful accessing UEs  decreases with the increase of the number of contenders~\cite{SUCRGBPA}. To improve the pilot resource utilization of the SUCRe scheme,  SUCR combined idle pilots access (SUCR-IPA) scheme was proposed in~\cite{SUCRIPA}, where the weaker  UEs randomly select idle pilots  to increase the number of successful accessing UEs.
  A user identity-aided pilot access scheme was proposed for massive MIMO with interleave-division multiple-access systems, where the interleaver of each UE is available at the BS according to the one-to-one correspondence between UE's identity number (ID) and its interleaver~\cite{IDPA}.  However, since the grant-based random access schemes require two handshake processes  between the base station (BS) and UEs, considering the small packet transmission in M2M communication,   such kind of  random access scheme will introduce heavy signaling overhead and low data transmission efficiency. To address this problem, the grant-free random access schemes have attracted much attention in recent years, which allow  active UEs to transmit their pilots and uplink messages  to the BS directly and performs activity detection, channel state information (CSI) estimation, and uplink message decoding in one shot.
%
 J. Ahn~\textit{et~al.}~proposed a Bayesian based random access scheme to detect  the UE's activity  and estimate the  CSI jointly by utilizing the expectation propagation algorithm, considering the  BS with one antenna~\cite{ce5}.  L. Liu~\textit{et~al.}~proposed a  approximate message passing (AMP) based grant-free scheme  to achieve the joint  activity detection and CSI estimation for massive MIMO systems~\cite{ce8}.
However,  this AMP-based grant-free  random access  scheme requires long pilot sequence to achieve better performance, resulting in heavy pilot overhead. These grant-free random access schemes  considers the single pilot structure, and Jiang~\textit{et~al.}~proposed to concatenate several multiple orthogonal sub-pilots into one pilot sequence, where different UEs are allocated different pilot sequences and the pilot sequence is utilized for activity detection and CSI estimation~\cite{MultiplePreambles}. The  performance comparison between these two kinds of 
pilot structures are made in~\cite{Preamble-Structures}.

%
%


The above-mentioned two kinds of random access schemes are based on the assumption that the BS has performed accurate time-frequency synchronization. However,  in practice, there are frequency errors caused by the Doppler shifts and/or frequency estimation errors  during the initial downlink synchronization,  and timing errors caused by the locations of UEs in the cell, which impair the pilot orthogonality and further degrade the access performance~\cite{tf1,tf2,tf4}. Considering the time-frequency asynchronous massive MIMO systems, L. Sanguinetti~\textit{et~al.}~proposed a random access scheme based on orthogonal frequency division multiplexing (OFDM) to solve the pilot collision by exploiting timing offsets and the large number of antennas~\cite{RATO}. However, the number of successful detected UEs is less than or equal to the number of subcarriers of the pilot, which cannot meet the massive access requirements of  the M2M communication.

To further resolve the pilot collision for the M2M communication in crowded asynchronous massive MIMO systems, we propose a novel random  access scheme,
where the BS  employs a proposed  estimation of signal parameters via rotational invariance technique enhanced (ESPRIT-E) method to estimate the effective timing offsets of UEs, and UEs judge whether it is detected in a distributed manner. Then, the detected UE  obtains its timing error from the effective  estimated timing error, and further compensates the timing error for uplink message transmission. Furthermore,  we analyze the   mean squared error (MSE) of the estimated effective timing offsets of UEs and the uplink throughput. Numerical results show  that, the proposed random access scheme significantly improves the uplink throughput,  and provide accurate value of the effective timing offset.

The remainder of this paper is organized as follows. System model is given in Section II. Section III describes the proposed random access process. We present the performance analysis in Section IV.  Simulation results and the conclusion are given in Section V and VI, respectively.

\textbf{Notation}: In this paper, we use the superscript `${\mathrm{T}}$', `*', and `${\mathrm{H}}$' to denote the transpose, complex conjugate, and conjugate transpose of a vector or a matrix, respectively. We use $\mathcal{C}\mathcal{N}( a,b )$ to  denote a circularly-symmetric complex Gaussian distribution with mean $a$ and variance $b$. Let $|| \cdot||$ indicate the Euclidean norm. We use $[\bm{x}]_n$  and $\bm{X}(i)$   to denote the $n^{th}$ element of vector $\bm{x}$ and  the $i^{th}$ column of matrix $\bm{X}$. Let `${\text{round}}$' denote the rounding operation. We utilize  $\text{arg(d)}$
to denote the phase of the complex $d$.


\section{SYSTEM MODEL}

 We consider the \mbox{time-division} duplexing (TDD) massive MIMO communication system based on OFDM. There are a BS with $M$ antennas located at the center of the cell and  $K$ single-antenna UEs uniformly distributed in the cell. We assume that each UE  becomes  active  with probability $p_a$. The number of UEs residing in the cell is $K$, and the number of active UEs is ${N_a}$.

The pilot with symbol length ${\tau}$
consists of $Q$  consecutive OFDM symbols in the time domain and $N$ adjacent subcarriers in the frequency domain~(i.e., ${\tau}=QN$). We use
 $\bm{{C}_{N}}=\{\bm{f}_0, \ldots, \bm{f}_i,  \ldots, \bm{f}_{N-1}\}$ ($\{{\bm{f}_i\in {\mathbb{C}}^{N}:\bm{f}_i^{\text{H}} \bm{f}_i ={N},~\forall{i}}\}$) and $\bm{{C}_{Q}}=\{\bm{t}_0, \ldots, \bm{t}_i,  \ldots, \bm{t}_{Q-1}\}$ ($\{{\bm{t}_i\in {\mathbb{C}}^{Q}:\bm{t}_i^{\text{H}} \bm{t}_i ={Q},~\forall{i}}$\}) to  represent the frequency domain code  set and time domain code set, respectively. The time domain code set $\bm{{C}_{Q}}$ can be any orthogonal sequence set, and the frequency domain code $\bm{f}_i$  is the Fourier basis, which is given by~\cite{RATO}
\begin{equation}\label{fi}
  [\bm{f}_i]_n=e^{j\frac{2\pi}{N}ni},~n=0,1,\ldots,N-1.
\end{equation}


 The received pilot signal of UE $k$ at the BS will introduce frequency error $w_k$ and timing error $\theta_k $.  Since the value of $w_k$ is very small in general and its impact can reasonably be neglected if the pilot
 contains only a few consecutive OFDM symbols~\cite{woffset,RATO}, we only consider the timing error $\theta_k=2D_k/(cT_s)$ where $D_k$ is the distance from UE $k$ to the BS, $c=3\times 10^8 m/s$ is the speed of light, $T_s=1/(\Delta f N_\text{FFT})$  is the sampling period where $N_\text{FFT}$ is the number of subcarriers with frequency spacing $\Delta f$. Note that, timing error $\theta_k$  appears as phase shifts at
the output of the receive discrete Fourier transform (DFT) unit~\cite{RATO}.

\section{The proposed random access scheme}
 \begin{figure}[h]
  \centering
  \includegraphics[scale=0.6]{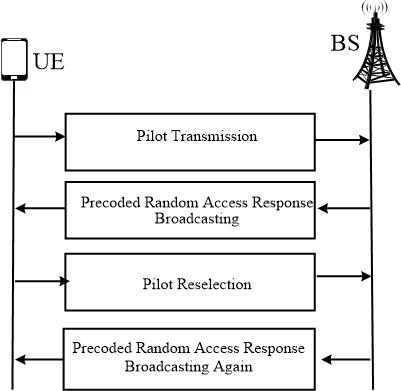}\\
  \caption{The proposed random access scheme.}\label{proposed scheme}
\end{figure}

Fig.~\ref{proposed scheme} shows the four steps of the proposed  random access scheme, and the details  are described as follows.


\subsection{Step 1: Pilot transmission}

Each UE randomly selects a frequency domain code from  set $\bm{{C}_{N}}$  and a time domain  code from set $\bm{{C}_{Q}}$. We use $l_k\in {(0, \ldots ,N-1)}$ and $i_k\in {(0, \ldots ,Q-1)}$ to represent the indexes of the frequency domain code and time  domain code selected by UE $k$, respectively. Thus, the pilot transmitted by UE $k$ is $\bm{f}_{l_k}\bm{t}_{i_k}^{\text{T}}$. Then, the received pilot signal at the $m^{th}$ antenna   over subcarrier $n$ is given by
\begin{equation}\label{0002}
 \begin{array}{rcl}
\bm{Y}_m(n) &=& \sum\limits_{k=1}^{{N_a}}{{\sqrt{\rho_k }}{h_k^m}{e^{-jn\theta _k\frac{2\pi}{N_\text{FFT}}} e^{-jnl_k\frac{2\pi}{N} } }{\bm{t}_{i_k}^{\text{T}}}} +\bm{W}_m^{n}\\
&=& \sum\limits_{k=1}^{{N_a}}{{\sqrt{\rho_k }}{h_k^m}{e^{-j2{\pi}n\epsilon _k}}{\bm{t}_{i_k}^{\text{T}}}} +\bm{W}_m^{n},
\end{array}
\end{equation}
where $\epsilon _k=\frac{l_k}{N}-\frac{\theta _k}{N_\text{FFT}}$ is the effective timing offset of UE $k$,  $\rho_k$ is the transmit power of UE $k$,  $\bm{W}_m^{n}\sim \mathcal{C}\mathcal{N}( 0,\sigma^2)$ is the additive noise and any future instances of the matrix or vector $\bm{W}$ with different sub- or superscripts will take the same distribution, and  $h_k^m$  denotes the channel response between UE $k$ and the $m^{th}$ antenna of the BS which is given by
\begin{equation}\label{2000}
              h_k^m = g_k^m\sqrt{\beta_k},
\end{equation}
 where $g_k^m\sim \text{ }\mathcal{C}\mathcal{N}(0,1)$ is the small-scale fading coefficient between UE $k$ and the $m^{th}$ antenna of the BS, $\beta_k$ is the channel gain of UE $k$ which is known to UE $k$. The channel between UE $k$ and the BS is denoted by $\bm{h}_k={(h_k^1,\ldots,h_k^M)^{\text{T}}}$. Then, the received pilot signal at the $m^{th}$ antenna of the BS can be written as
\begin{equation}\label{0003}
 \begin{array}{rcl}
\bm{Y}_m &=& \sum\limits_{k=1}^{{N_a}}{{\sqrt{\rho_k }}{h_k^m}{([1,\cdots,e^{j2\pi (N-1)\epsilon _k}])}{\bm{t}_{i_k}^{\text{T}}}} +\bm{N}_m\\
&=&\sum\limits_{k=1}^{{N_a}}{{\sqrt{\rho_k\beta_k }}{g_k^m}{\bm{c}(\epsilon _k)}{\bm{t}_{i_k}^{\text{T}}}} +\bm{N}_m,
 \end{array}
\end{equation}
where ${\bm{c}(\epsilon _k)}=[1,\cdots, e^{j2\pi (N-1)\epsilon _k}]$ stands for the effective frequency domain code of UE $k$, and $j$ is the unit
imaginary number. Furthermore, throughout this paper, we consider that ${{\rho_k}{{\beta }_{k}}} = 1$, which can be achieved by the power control mechanism \cite{POWER}. This ensures that the received signals from UEs  have the same power, and thus obtains a fair estimation performance.

\subsection{Step 2: Precoded random access broadcasting}

Based on the received pilot signal at the $m^{th}$ antenna of the BS $\bm{Y}_m$, the BS  sends the precoded random access response to active UEs. The procedure is described as follows.

\begin{enumerate}
  \item The number of active UEs estimation

$\bm{Y}_m$ is first correlated with the time domain code $\bm{t_i}$ $(i=0,\ldots,Q-1)$ in set $\bm{{C}_{Q}}$,
\begin{equation}\label{2004}
\bm{z}_m^i = {\bm{Y}_m}\frac{\bm{t_i}^*}{||\bm{t_i}||}=\sum\limits_{u\in \bm{\mathcal{A}_i}}{{g_k^m}([1,\ldots,e^{j2\pi (N-1)\epsilon _u}])}+{\bm{W}_m^{'}},
\end{equation}
where $\bm{\mathcal{A}_i}$ is the set of UEs selecting $\bm{t_i}$, and  ${\bm{W}_m^{'}}={\bm{W}_m}{\bm{t_i}^*}/{||\bm{t_i}||}$.

Let $\bm{Z}^{i}=[\bm{z}_1^i,\bm{z}_2^i,\ldots,\bm{z}_M^i]^\text{T}$. Then, by correlating  the first column in $\bm{Z}^i$ with the matrix $\bm{Z}^i$,  we have
\begin{equation}\label{11}
\bm{z}_c^i=  { \frac{{\bm{Z}^i(1)}^{\text{H}}{\bm{Z}^i}}{M} } \\
          \xrightarrow{M \to +\infty }\sum\limits_{u\in \bm{\mathcal{A}_i}}{([1,\ldots,e^{j2\pi (N-1)\epsilon_u}])}.
\end{equation}
Eq.~({\ref{11}}) is obtained based on the propagation of the  massive MIMO channel~\cite{propagation}, i.e.,
\begin{equation}\label{2001}
\begin{aligned}
&  \lim_{{M} \to +\infty}{ \frac{ {\bm{g}_p}^{\text{H}} {\bm{g}_u} }{M} }= {0},     p\neq u,  \\
&  \lim_{{M} \to +\infty}{ \frac{  {\bm{g}_p}^{\text{H}} {\bm{g}_p} }{M} } ={1}.
\end{aligned}
\end{equation}

Based on Eq.~({\ref{11}}), we observe that, when $M$ goes into infinity,  the $n^{th}$ element  in  $\bm{z}_c^i$  is indeed the sum of the effective frequency domain code over the $n^{th}$ subcarrier of active UEs selecting  the time domain code $t_i$. Actually, the first element  in  $\bm{z}_c^i$ is the number of active UEs selecting  the time domain code $t_i$. Therefore, we can utilize the first element in $\bm{z}_c^i$ (i.e., $[\bm{z}_c^i]_1$)  to estimate the number of UEs selecting the time domain code $t_i$, denoted by $\overline{N_a^{i}}$.

\item The effective timing offsets  estimation

By utilizing $\overline{N_a^{i}}$, we utilize the proposed ESPRIT-E method to estimate  the  effective timing offsets of UEs, which is described as follows.

The sample covariance matrix ${\bm{R_z^i}}$ associated to $\bm{z}_m^i$ is computed by
\begin{equation}\label{2006}
 {\bm{R_z^i}}=\frac{1}{M}\sum\limits_{m=1}^{M}{\bm{z}_m^i}{({\bm{z}_m^i})^{\text{H}}}.
\end{equation}
By utilizing the eigenvectors of $\bm{R_z^i}$  associated to $d^{i}=\min(N-1,\overline{N_a^{i}})$ largest eigenvalues in $\bm{R_z^i}$ to form a new matrix $\bm{V^i}$, the ESPRIT method is utilized to estimate the  effective timing offset of the $s^{th}$ UE selecting the time domain code $t_i$~\cite{ESPRIT}, denoted by $\overline{\epsilon_{I_i^s}}$
\begin{equation}\label{2008}
           \overline{\epsilon_{I_i^s}}=\frac{\arg \{ \psi_s^i \} }{2\pi}, \quad s=1,2,\ldots,{d^{i}},
           \end{equation}
where $I_i^s$ is the index of UE among the $N_a$ active UEs, $\{ \psi_1^i,\psi_2^i,\cdots,\psi_2^{d^{i}} \}$ are  eigenvalues of matrix $({{{\bm{V}_1^i}^{\text{H}}{\bm{V}_1^i} })^{-1}}{{\bm{V}_1^i}^{\text{H}}{\bm{V}_2^i}}$, and the matrices $\bm{V}_1^i$ and $\bm{V}_2^i$ are obtained by taking the first and the last $N-1$ rows of $\bm{V^i}$, respectively.

The  ESPRIT method can only estimate the effective timing offsets of ${d^{i}}\leq (N-1$)  UEs. If the value of $\overline{N_a^{i}}$ is larger than $(N-1)$,
Based on Eq.~(\ref{11}), after subtracting these $d^{i}$ estimated  effective timing offsets  from  $\bm{z}_c^i$, we can obtain the sum of the effective timing offsets of the remaining $(\overline{N_a^{i}}-d^{i})$ UEs, which is given by
\begin{equation}\label{2009}
\begin{aligned}
           \sum\limits_{u=d^{i}+1}^{\overline{N_a^{i}}}e^{j2\pi (1-1)\epsilon_{I_i^u}}&= \bm{z}_c^i(1)-\sum\limits_{s=1}^{{d^{i}}}e^{j2\pi (1-1)\overline{\epsilon_{I_i^s}}},\\
            \sum\limits_{u=d^{i}+1}^{\overline{N_a^{i}}}e^{j2\pi (2-1)\epsilon_{I_i^u}}&= \bm{z}_c^i(2)-\sum\limits_{s=1}^{{d^{i}}}e^{j2\pi (2-1)\overline{\epsilon_{I_i^s}}},\\
           &\cdots \\
           \sum\limits_{u=d^{i}+1}^{\overline{N_a^{i}}}e^{j2\pi (N-1)\epsilon_{I_i^u}}&= \bm{z}_c^i(N)-\sum\limits_{s=1}^{{d^{i}}}e^{j2\pi (N-1)\overline{\epsilon_{I_i^s}}}.
\end{aligned}
 \end{equation}
By solving Eqs.~(\ref{2009}), we can obtain the effective timing offsets of the remaining ($\overline{N_a^{i}}-d^{i}$) UEs. Thus, the estimated effective timing offsets  are $\{\overline{\epsilon_{I_i^{1}}},\ldots, \overline{\epsilon_{I_i^{s}}},\ldots,\overline{\epsilon_{I_i^{{\overline{N_a^{i}}}}}}\}$.

\item{  Channel response estimation}

By utilizing $\overline{\epsilon_{I_i^{s}}}$, we employ the least squares (LS) method to  estimate the channel response of UE ${I_i^s}$ between UE ${I_i^s}$ and  the $m^{th}$ antenna at the BS
\begin{equation}\label{2010}
           {h_{I_i^s}^m}=({{\bm{c}(\overline{\epsilon_{I_i^{s}}})}}^{\text{H}}{{\bm{c}(\overline{\epsilon _{I_i^{s}}})}} )^{-1}{{\bm{c}(\overline{\epsilon_{I_i^{s}}})}}^{\text{H}}{\bm{z}_m^i}, \quad s=1,2,\ldots,\overline{N_a^{i}}
            \end{equation}
\end{enumerate}

Based on procedures 1)-3), the BS can obtain the estimated  channel responses of UEs selecting other time domain codes. The BS broadcasts the precoded random access response ${{\bm{h}_{I_i^s}}{\bm{t_i}}^{\text{T}}},~(i=0,\ldots,Q-1, ~ s=1,\ldots,{\overline{N_a^{i}}})$ and the corresponding effective timing error $\overline{\epsilon_{I_i^{s}}}$ to all active UEs.

\subsection{Step 3: Pilot reselection}

The received signal ${\bm{R}_k^{i,j}}$   at UE $k$  is written as
 \begin{equation}\label{2002}
              {\bm{R}_k^{i,s}} =  { {\bm{h}_k}^{\text{H}} {{\bm{h}_{I_i^s}}{\bm{t_i}}^{\text{T}}} }
              + {\bm{W}_k}, i=0,\ldots,Q-1, ~ s=1,\ldots,{\overline{N_a^{i}}}.
 \end{equation}
UE $k$ first correlates the received signal ${\bm{R}_k^{i,s}}$ with its selected
time domain code to obtain
\begin{equation}\label{3001}
\begin{aligned}
{r_k^{i,s}}&={{\bm{R}_k^{i,s}}} \frac{{\bm{t_{i_k}}}^*}{||{\bm{t_{i_k}}}||M\sqrt{\beta_k}}\\
&=\frac{ {\bm{h}_k}^{\text{H}} {{\bm{h}_{I_i^s}}}}{M\sqrt{\beta_k}}+{{\bm{W}_k}} \frac{{\bm{t_{i_k}}}^*}{||{\bm{t_{i_k}}}||M\sqrt{\beta_k}},~i=i_k,~j=1,\ldots,\overline{N_a^{i}}.\\ \\
&\xrightarrow{M \to \infty } \left\{ \begin{array}{l}
1, \text{if}~~ {\bm{h}_{I_i^s}}\approx {\bm{h}_k},\\
0, \text{otherwise}.
\end{array} \right.
\end{aligned}
\end{equation}

Then, UE $k$ uses the following rule to judge whether it is detected in a distributed manner, which is given by
\begin{equation}\label{3002}
\begin{aligned}
&{D_k}:\text{if}~\sum\limits_{j=1}^{\overline{N_a^{i}}}
\text{round}({r_k^{i,s}})=1  &&(\text{Detected})  \\
&{U_k}:\text{otherwise}              &&(\text{Undetected})
\end{aligned}
\end{equation}

If UE $k$ is a detected UE, it uses the effective timing error $\overline{\epsilon_{I_i^j}}$ (i.e., $I_i^s=k$)~corresponding to  ${\bm{h}_{I_i^s}}$ that makes $\text{round}({r_k^{i,s}})=1$ to obtain its timing error as follows
\begin{equation}\label{8090}
  \overline{\theta_{k}}=N_\text{FFT}(  \frac{l_{k}}{N} - \overline{\epsilon_{I_i^s}}).
\end{equation}
Then, UE $k$ employs  $\overline{\theta_{k}}$ to compensate its timing
error for uplink message transmission. Otherwise, UE $k$  randomly reselects a frequency domain code from  set $\bm{{C}_{N}}$  and a time domain  code from set $\bm{{C}_{Q}}$ to obtain its pilot, and send it to the BS, as we described in step 1.

\subsection{Step 4: Precoded random access response broadcasting again}

Similar to step 2, based on the received  pilot signal,  the BS generates and broadcasts the  precoded random access responses again,  and each UE  reselecting its pilot during step 3 employs the rule in Eq.~(\ref{3002}) to judge whether it is detected. If UE $p$ is a detected UE, it compensates its timing error for uplink message transmission. In the following, all detected UEs send their uplink messages to the BS, and thus the BS can
utilize any blind detection method to obtain  their uplink messages, such as the proposed EICA method proposed in~\cite{EICA}, which is not the focus of this paper.

 \begin{remark}[Why we utilize Eq.~({\ref{3002}}) as the detection rule]
 Eq.~(\ref{3001}) indicates that, if the estimated channel response ${\bm{h}_{I_i^s}}$  is approximately equal to the channel response of UE $k$, the value of ${r_k^{i,s}}$ equals $1$ with large value of $M$. Furthermore, based on Eq.~(\ref{2010}), we observe that multiple  similar  effective timing offsets lead  to multiple estimated channel responses being approximately equal to the channel response of UE $k$, resulting in $\sum\limits_{j=1}^{\overline{N_a^{i}}}\text{round}({r_k^{i,s}})>1$. However, we cannot determine the timing error of UE $k$ for such case, because of $\epsilon_k=\frac{l_k}{N}-\frac{\theta _k}{N_\text{FFT}}$ which means that UEs with different selected frequency domain codes and different timing errors may have similar  effective timing offsets. To obtain the timing error of UE $k$, if multiple estimated channel responses are approximately equal to the channel response of UE $k$, i.e.,  $\sum\limits_{j=1}^{\overline{N_a^{i}}}\text{round}({r_k^{i,s}})>1$, we claim that  UE $k$ is not detected. Obviously, the case of  $\sum\limits_{s=1}^{\overline{N_a^{i}}}\text{round}({r_k^{i,s}})=0$ means that UE $k$ is not detected.
The case of  $\sum\limits_{s=1}^{\overline{N_a^{i}}}\text{round}({r_k^{i,s}})=1$ indicates that there are no similar effective timing offsets with UE $k$, and thus we can obtain its timing error based on Eq.~({\ref{8090}}).
\end{remark}

\section{Performance analysis}
In this section, we analyze the performance of the proposed random access scheme, including the MSE of the estimated effective timing offset, and the uplink throughput.

\subsection{MSE of the estimated effective timing offset}
In the proposed random access scheme, we utilize the  ESPRIT-E method to estimate  the  effective timing offsets of UEs. Specifically, for the active UEs selecting the same time domain code, we first utilize the ESPRIT method to estimate the effective timing offsets of $N-1$ active  UEs. Then, after subtracting the $N-1$  estimated effective timing offsets from the sum of effective timing offsets of all active UEs, we obtain the  effective timing offsets of the remaining active UEs by solving the  polynomial equations in Eqs.~(\ref{2009}). Obviously, the procedure of Eqs.~(\ref{2009}) will introduce  noise.
Furthermore, based on Eq.~(\ref{11}), the sum of effective timing offsets of all active UEs selecting the same time domain code is accurate when $M$ goes to infinity. Hence, the MSE of the estimated effective timing offset  of the proposed  ESPRIT-E method is greater than or equal to  that of the estimated effective timing offset when $M$ goes to infinity, which is given by
\begin{equation}\label{4001}
\begin{aligned}
\text{MSE}&=\frac{1}{{{N_s}}}\sum\limits_{k = 1}^{{N_s}} {({\theta _k} - \overline{\theta _k})}^2 \\
&\ge  \text{MSE}_{M \to \infty},
\end{aligned}
\end{equation}
where ${N_s}$ is the number of UEs being detected. We utilize the Monte Carlo simulation
method to obtain the value of $\text{MSE}_{M \to \infty}$, which is the lower bound of the MSE of the estimated effective timing offset  of the ESPRIT-E method.


\subsection{Uplink throughput analysis}

We define the uplink throughput as the number of the successful detected active UEs. When given the number of active UEs selecting the same time domain code, based on Eqs.~(\ref{2009}), the  larger the value of $M$, the more accurate the estimated effective timing offsets. This leads to the increase of the number of the detected active UEs. Hence, we can obtain the upper bound of the uplink throughput of the proposed random access protocol by setting the value of $M$ to infinity. Since it is hard to derive the analysis results of the uplink throughput for any values of $M$, we utilize the Monte Carlo simulation method  to obtain the upper bound, denoted by $T_{u}$.

\section{Simulation results}

In this section, we compare the performance of the proposed random access scheme with the random access scheme in~\cite{RATO}, including the  MSE of the estimated effective timing offset and the uplink throughput. In addition, terms~``the proposed random access scheme"  and ``the random access scheme in~\cite{RATO}" are abbreviated as ``The proposed RA" and ``RA in~\cite{RATO}" in the result figures, respectively.

\begin{table}[h]
\caption{System Parameters}
\label{table-p}
\setlength{\tabcolsep}{3pt}
\begin{tabular}{|p{100pt}|p{120pt}|}
\hline
Parameter&
Value\\
\hline
layout &
Regular hexagonal cell\\
Cell radius &
250m \\
The number of UEs&
$N_a$=6-22 \\
Number of antennas & $
M=$20-400 \\
Bandwidth &
$B$=20MHz \\
DFT size &
$N_\text{FFT}$=1024 \\
The number of time-domain codes &
$Q$=2 \\
The number of frequency-domain codes &
$N$=8,12 \\
\hline
\end{tabular}
\label{tab1}
\end{table}

In the simulation, we consider a cellular network operating over a bandwidth $B$=20 MHz and the radius of the cell is 250 meters and all UEs locate uniformly at the place farther than 25 meters from the BS. Table~\ref{table-p} shows the simulation parameters setting.


 \begin{figure}[h]
  \centering
  \includegraphics[height=6cm,width=8.5cm]{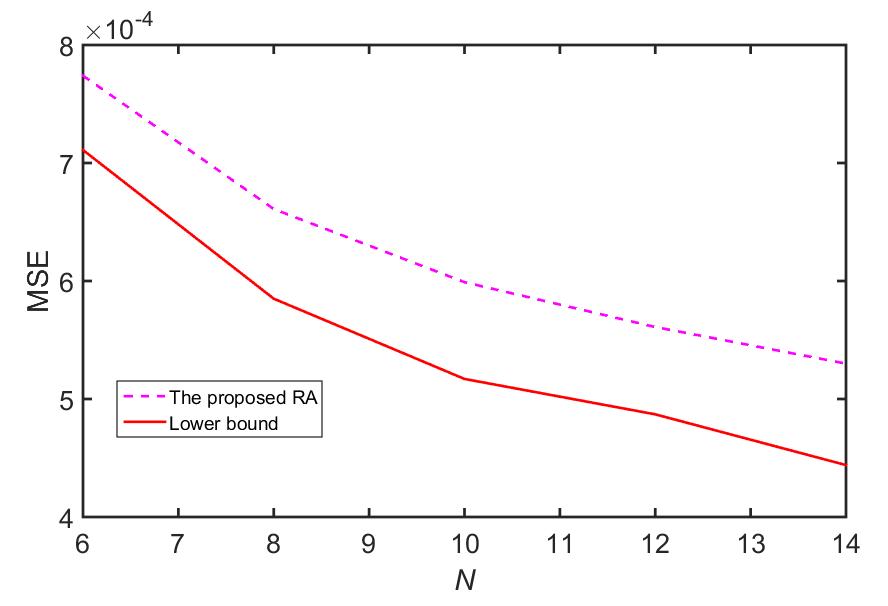}\\
  \caption{MSE versus the number of subcarriers.}\label{MSE-N}
\end{figure}

 \begin{figure}[h]
  \centering
  \includegraphics[height=6cm,width=8.5cm]{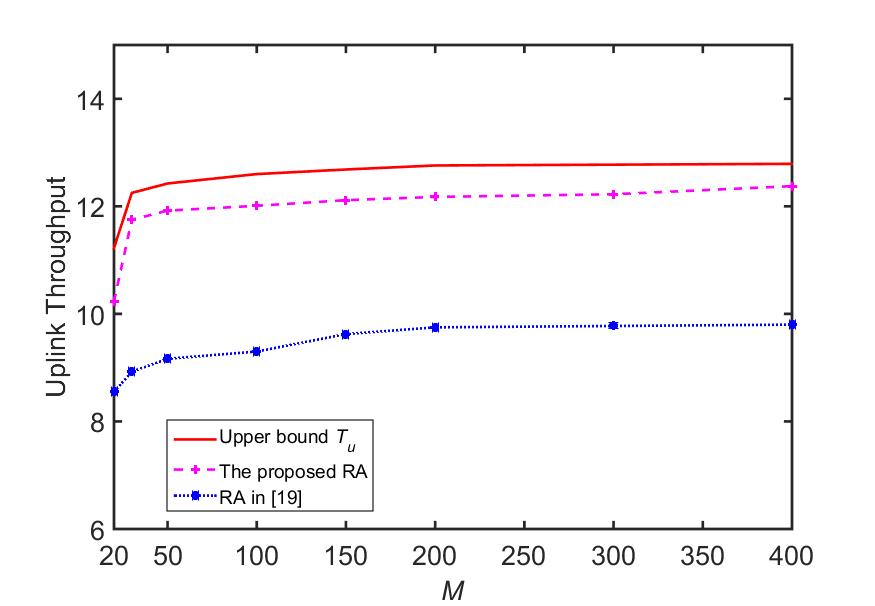}\\
  \caption{Uplink throughput versus the number of antennas at the BS.}\label{UT-M}
\end{figure}

 \begin{figure}[h]
  \centering
  \includegraphics[height=6cm,width=8.5cm]{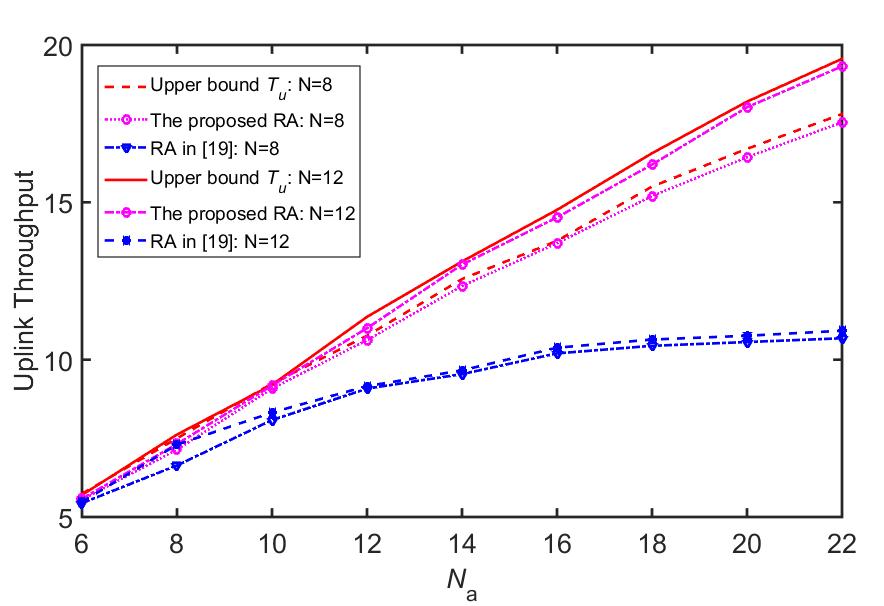}\\
  \caption{Uplink throughput versus the number of active UEs.}\label{UT-NA}
\end{figure}

Fig.\ref{MSE-N} shows how the MSE of the estimated effective timing offset  changes with  the number of subcarriers under $Q=2, M=200$, and $N_a=6$, to verify the proposed effective timing offset estimation method.  The simulation results show that the MSE of the  estimated effective timing offset takes small value, and decreases with  the number of subcarriers $N$. The reason is that, the increase of the number of subcarriers $N$, means the increase of the number of pilots, resulting in the decrease of the pilot collision probability and the interference between UEs. We can also note that, the MSE of the  estimated effective timing offset  is close to the lower bound.

Fig.\ref{UT-M} shows how the uplink throughput changes with the number of antennas at the BS under $Q=2, N=8$, and $N_a=14$. We can observe from the simulation results  that the uplink throughputs of the proposed random access scheme and the random access scheme in~\cite{RATO} increase  dramatically from $M = 20$ to $M = 50$, and increases at a slower pace when $M \ge 50$. We also note that, the uplink throughput of the proposed random access is significantly higher than that of the random access scheme in~\cite{RATO}, and much close to the upper bound $T_{u}$.  The reason is that,  the random access scheme in~\cite{RATO} utilizes the ESPRIT method to estimate the effective timing offsets, and thus the number of detected effective timing offsets is limited by the number of subcarriers $N$. However, to address this problem, our proposed random access scheme proposed  an ESPRIT-enhanced (i.e., ESPRIT-E) method to estimated the effective timing offsets.

Fig.\ref{UT-NA} shows how the uplink throughput changes with the number of active UEs under $Q=2, M=200$, and $N=8,12$. We can observe from the simulation results  that the uplink throughput of the proposed random access scheme  is significantly higher than  the random access scheme in~\cite{RATO} with the increase of the number of active UEs, and close to the upper bound $T_{u}$. We also see that, with the increase of the number of active UEs, the uplink throughput of the proposed random access scheme increases almost linearly, whereas that of the random access scheme in~\cite{RATO} increase almost linearly from $N_a = 6$ to $N_a= 8$, and increases at a slower pace when $N_a \ge 8$. The reason is the same as we described for Fig.\ref{UT-M}, i.e., the number of detected UEs is limited by the number of subcarriers $N$.

\section{Conclusion}
In this paper, we proposed a new random access scheme for M2M communication in crowded asynchronous massive MIMO systems to resolve the intra-cell pilot collision. The proposed random access scheme estimates the effective timing offsets by utilizing the proposed ESPRIT-E method, and then the UE can obtain its timing errors for uplink message transmission. 
We also analyze the performance of the proposed random access scheme, including the  MSE of the estimated effective timing offset and the uplink throughput. 
Simulation results show that, 
compared to the exiting random access scheme for the crowded asynchronous massive MIMO systems, 
the proposed random access scheme can improve the uplink throughput and provide  accurate effective timing offsets at the same time.

\bibliographystyle{IEEEtran}
\bibliography{ref}


\end{document}